\documentclass[pre,amsmath, twocolumn, superscriptaddress,showpacs]{revtex4-1}
\usepackage{graphicx}
\usepackage{color}
\usepackage{latexsym}

\begin{document}

\title{Minimum energetic cost to maintain a target nonequilibrium state}

\author{Jordan M. Horowitz}
\author{Kevin Zhou}
\author{Jeremy L. England}
\affiliation{Physics of Living Systems Group, Department of Physics, Massachusetts Institute of Technology, 400 Technology Square, Cambridge, MA 02139}

\date{\today}

\begin{abstract}  
In the absence of external driving, a system exposed to thermal fluctuations will relax to equilibrium.  However, the constant input of work makes it possible to counteract this relaxation, and maintain the system in a nonequilibrium steady state.  In this Article, we use the stochastic thermodynamics of Markov jump processes to compute the minimum rate at which energy must be supplied and dissipated to maintain an arbitrary nonequilibrium distribution in a given energy landscape.  This lower bound  depends on two factors: the undriven probability current in the equilibrium state, and the distance from thermal equilibrium of the target distribution.  By showing the consequences of this result in a few simple examples, we suggest general implications for the required energetic costs of macromolecular repair and cytosolic protein localization. 
\end{abstract}

\maketitle 

\section{Introduction}

In many functional contexts -- nano-engineering and bio-molecular assembly, to name a few -- it is essential to be able to maintain a system in a nonequilibrium steady state.  Thermal and chemical equilibria are generally dominated by configurations with low energy and high internal entropy, yet there are many situations in which the useful outcome is either highly ordered, high in energy, or both. For example, equilibrium protein solutions misfold and aggregate irreversibly at concentrations comparable to those found in the cell. To avoid this, cells continually harness chemical work by consuming ATP to fuel the molecular chaperones that hold back aggregation~\cite{luby1999cytoarchitecture,kim2013molecular}.

The preceding observations point to a clear question of general importance: what is the minimum rate of energy input required to maintain a desired nonequilibrium distribution over states of known, fixed energies? 
While previous studies determined the minimum energy required to isothermally prepare a system in an arbitrary nonequilibrium distribution -- or conversely the maximum work extractable from relaxation to equilibrium --\cite{Procaccia1976,Kawai2007,Hasegawa2010,Takara2010,Deffner2011,Esposito2011}, the power cost required to hold a system in a desired nonequilibrium distribution is a distinct and significant thermodynamic quantity that has not been analyzed.
In this Article, we employ the tools of modern nonequilibrium thermodynamics~\cite{Seifert2012,VandenBroeck2015} to compute this cost and show that it is fully determined by two factors: first, an information-theoretic measure of the driven steady state's distance from equilibrium, and second, the magnitude of probability flux in the system's undriven equilibrium state.  

\section{Driven repair in a two-level system} 
\begin{figure}[t]
\includegraphics[scale=.48]{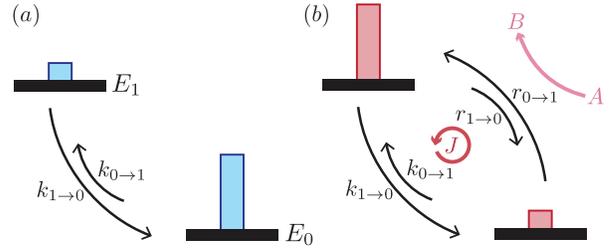}
\caption{Two-state system schematic: (a) Undriven two-state system with energies $E_0 < E_1$ and equilibrium probability distribution pictured as uneven blue rectangles. (b) Driven nonequilibrium state with stabilized target configuration $1$, supported by the additional driven pathway powered by chemical work due to coincident conversion of chemical species $A\to B$.
}
\label{fig:two}
\end{figure}

Before addressing the general theory, we first consider an illustrative example that captures the same essential physics.  Consider the two-level system depicted in Fig.~\ref{fig:two},  
which makes stochastic transitions between two distinct microstates. To be concrete, these two states could represent, on the one hand, a mono-dispersed arrangement of two native, functional macromolecules, and on the other hand, an inactive, aggregated dimer. The first of these states (labeled $1$ with energy $E_{1}$) is desired for its functionality, so we want it to be favored; the latter state (labeled $0$ with energy $E_{0}$ such that $\beta\Delta E=\beta(E_{1}-E_{0}) > 1 $), is an unwanted aberration that we would like to suppress by driving disaggregation. 
Prior to control, the system is governed by Poissonian transition rates $k_{1\to 0}$ and $k_{0\to  1}$ that preserve the Boltzmann distribution $p^{\rm eq}_{1}/p^{\rm eq}_{0}=k_{0\to  1}/k_{1\to 0}=e^{-\beta(E_{1}-E_{0})}$~\cite{VanKampen}, which we assume favors the unwanted state.  
The control goal is then to maintain the system in a specified nonequilibrium distribution obeying the occupancy ratio $\rho = p^{\rm neq}_{1}/p^{\rm neq}_{0}>e^{-\beta(E_1-E_0)}$.   

There are often multiple such jump-type reactions that connect a pair of states, even at equilibrium.  
For example, one process moving from $1$ to $0$ could be the dissociation of a pair of dimerized protein monomers solely through thermal fluctuation, and have probability rate $a_{0\to 1}^{(1)}$.  
Another, with rate $a_{0\to 1}^{(2)}$, might be a transition that occurs via the normal pathway of an assisting molecular chaperone, but, crucially, \emph{without} the hydrolysis of ATP.  The latter often is so unlikely that it is typically neglected in a biophysical model.
However, for reasons of thermodynamic consistency, we must allow it to be possible in principle.  Thus, the total undriven probability rate of dimerization would be $k_{1\to 0} = \sum_{l}a^{(l)}_{1\to 0}$.  

The cell then often implements control by populating the target state through a collection of driven auxiliary transition pathways that consume energy from an ambient source, such as the hydrolysis of ATP. 
For example, protein folding and aggregation is managed \emph{in vivo} by the activity of molecular chaperone ATPases ~\cite{desantis2012operational,kim2013molecular}.
Thus, these controlled transitions usually only become relevant once an external drive (such as chemical baths of ATP and ADP) is introduced, allowing the execution of the same motion as the undriven dissociation event while an ATP happens to be hydrolyzed, which is a physically distinct pathway from the dissociation event mediated by the chaperone without ATP hydrolysis.

In light of this discussion, we are motivated to implement control through the addition of a supplementary transition pathway to our two-state toy model driven by a thermodynamic force; a chemical example would be to link the transition to a chemical reaction $A\to B$ down a chemical potential gradient $\Delta\mu=\mu_{A}-\mu_{B}>0$~\cite{Seifert2012,Parrondo2002}. 
For the dynamics to be thermodynamically consistent, the rates around the induced cycle due to the inclusion of $r_{1\to 0}$ and $r_{0\to  1}$ must match the thermodynamic force around the cycle via~\cite{Seifert2012,Parrondo2002,Qian2005b}
\begin{equation}\label{eq:db2}
\beta\Delta\mu=\ln\left(\frac{k_{1\to 0}}{k_{0\to1}}\frac{r_{0\to 1}}{r_{1\to 0}}\right).
\end{equation}

With the driving, the system relaxes to a nonequilibrium steady state, accompanied by a continual probability flux $J=p_1^{\rm neq}k_{1\to 0}-p_0^{\rm neq}k_{0\to1}$ as the system preferentially flows down the equilibrium pathway and back up the driven pathway, see Fig.~\ref{fig:two}. This cycle is maintained by the chemical potential gradient $\Delta\mu$ that does chemical work at a rate~\cite{Parrondo2002}
 \begin{equation}\label{eq:Wchem}
\beta{\dot W}_{\rm chem}= J\cdot\beta\Delta\mu=J\ln\left(\frac{k_{1\to 0}}{k_{0\to1}}\frac{r_{0\to1}}{r_{1\to 0}}\right),
 \end{equation}
 which quantifies the energetic cost to maintain the nonequilibrium state.

Our objective is to minimize the steady-state energy consumption ${\dot W}_{\rm chem}$ at fixed $\rho$. From \eqref{eq:Wchem}, we see that ${\dot W}_{\rm chem}$ splits into two additive terms, each weighted by the current $J$. The first, proportional to $\ln(k_{1\to 0}/k_{0\to1})=\beta\Delta E$, is independent of how we drive the system, simply reflecting properties of the undriven kinetics. The second, proportional to $\ln(r_{0\to1}/r_{1\to 0})$, we can vary. We can make progress on finding the optimal ratio by first considering the special case of extremely fast driven transitions: $k/r\to 0$. In this case, to maintain the nonequilibrium ratio $\rho$, we must have $r_{0\to1}/r_{1\to 0}=\rho$, since the desired steady state is entirely determined by the probability flow back and forth over the fast, driven transition. Away from this limit, when $k/r>0$, the relative effect of the undriven transitions on the dynamics is enhanced.  This effect must be compensated by a stronger asymmetry of the driven transition rates to maintain $\rho$, which means $r_{0\to1}/r_{1\to 0}>\rho$. Looking back to ${\dot W}_{\rm chem}$ \eqref{eq:Wchem}, we see this requires a higher rate of dissipation than the optimal $k/r\to 0$. Thus, the minimum cost is
\begin{equation}\label{eq:WChemMin}
\beta{\dot W}_{\rm chem}\ge J\ln\left(\frac{k_{1\to 0}}{k_{0\to1}}\frac{p^{\rm neq}_{1}}{p^{\rm neq}_{0}}\right)=J\left(\ln\rho+\beta\Delta E\right).
\end{equation}

In this simple model, the minimum work depends on three things: the energy gap $\Delta E$, the imbalance of probability $\rho$ in the target distribution, and the current $J$ determined by the transition rate $k\equiv k_{1\to 0}$ out of the target state (as $k_{0\to1}=ke^{-\beta\Delta E}$).
With these parameters \eqref{eq:WChemMin} takes the illuminating form
 \begin{equation}
\beta {\dot W}_{\rm chem} \ge k\frac{\rho}{1+\rho}\left(1-\frac{1}{\rho e^{\beta\Delta E}}\right)\left[\ln\rho+\beta\Delta E\right].  
 \end{equation}
Observe that the basic timescale is the undriven transition rate $k$ at which thermal fluctuations cause spontaneous transitions from the desired state to the aberrant ``damaged'' state.  
The last factor, in brackets, dictates the remaining physics and exhibits two distinct regimes:
For $\ln\rho \approx \beta\Delta E$, the undriven energy difference makes a significant contribution to the minimum cost.  However, as our demand for fidelity increases, $\ln\rho\gg \beta \Delta E$, the determining factor is our fidelity criterion $\rho$, which captures how strongly we pump into the target state.

Although the preceding remarks were specific to one simple system, the physics behind them is general.  In what follows, we  provide a proof of this general lower bound on the rate of dissipation for an arbitrarily driven Markov jump process.

\section{Setup}
\begin{figure}[tb]
\includegraphics[scale=.225]{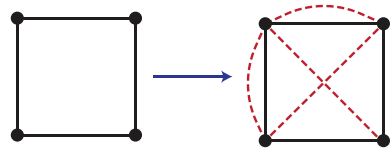}
\caption{Illustration of Markov jump process state graph: Nodes represent mesostates and edges allowed transitions.  Control is implemented by adding transitions (red dashed edges) that push the system into a desired nonequilibrium steady-state distribution $p^*\neq p^{\rm eq}$.}
\label{fig:graph}
\end{figure}

 Consider a system making stochastic transitions among a set of discrete mesostates, or configurations, $i=1,\dots, N$, with (free) energies $E_i$.
We can visualize these dynamics occurring on a graph like in Fig.~\ref{fig:graph}, where each configuration is assigned a node, and possible transitions are represented by edges (or links).

The dynamics are modeled as a Markov jump process according to transition rates $R_{ij}$ from $j$ to $i$, with $R_{ij}\neq 0$ only when $R_{ji}\neq0$. As such, the system's time-dependent probability distribution $p_i(t)$ evolves according to the Master equation~\cite{VanKampen}
\begin{equation}\label{eq:master}
\partial_t p_i(t) = \sum_{j\neq i} R_{ij}p_j(t)-R_{ji}p_i(t) \equiv\sum_{j\neq i} J_{ij}(p),
\end{equation}
 with probability currents $J_{ij}(p)$.

In the absence of any control, we assume that our system relaxes to a thermal equilibrium steady state at inverse temperature $\beta=1/T$, given by the Boltzmann distribution $p_i^{\rm eq}=e^{\beta(F^{\rm eq}-E_i)}$ with equilibrium free energy $F^{\rm eq}=-T\ln\sum_i e^{-\beta E_i}$; where from here on we set Boltzmann's constant to unity, $k_{\rm B}=1$.
To guarantee equilibrium, we impose detailed balance on the transition rates~\cite{VanKampen}
\begin{equation}\label{eq:db}
R_{ij}p_j^{\rm eq}=R_{ji}p_i^{\rm eq}.
\end{equation}
In equilibrium, each transition is balanced by its reverse.
Our goal is to maintain the system in a target nonequilibrium steady state $p^*\neq p^{\rm eq}$ and to calculate the minimum dissipation required.

\section{Minimum dissipation cost} 

When discussing a minimum energetic cost, it is first necessary to specify the set of allowable controls. The most comprehensive set would be complete control over the the system's energies $\{E_i\}$. We could then fix the system in $p^*$ by shifting all the energies to $E^*_i=-T\ln p^*_i$, thereby making the target state $p^*$ the new equilibrium. While there is a one-time energetic cost to change the energies (namely, the nonequilibrium free energy difference)~\cite{Esposito2011}; afterwards the system is maintained in $p^*$ for free.
However, cells frequently do not utilize this mechanism; in numerous biochemical examples, free energies of states remain fixed, and structural fidelity is achieved by coupling various dissipative processes.
For example, the free energy difference between a folded and unfolded protein sets the baseline rate of undriven thermal transitions, and then a distinct driven transition pathway mediated by molecular chaperones is added to shift the relative stability of the protein's configurations~\cite{jiang2014reca}. 

Motivated by this observation, we take the energies $\{E_i\}$ to be static parameters fixing the thermal transition rates and modify the steady-state distribution by introducing additional ``control" transitions with transition rates $\{M_{kl}\}$, as in Fig.~\ref{fig:graph}.
As was outlined in our analysis of the two-state system at the beginning of this article, we assume that for every molecular reaction contributing to the total probability of an undriven edge in the Markov graph, there is a corresponding process contributing to the driven (``control") edge that is accompanied by exchange with one or more external baths.  This requirement is general to any physically consistent description of matter coupled to heat and chemical baths, though it often can be safely ignored since many of the contributing processes are so unlikely that they contribute nothing to the physics.  Since we are modeling the thermodynamics of the general case, however, it is appropriate to point out this pairing between driven and undriven transitions.

Our only additional assumption is that the rates $\{M_{kl}\}$ satisfy a local detailed balance relation,
\begin{equation}\label{eq:dbM}
\ln\frac{M_{kl}}{M_{lk}}=\Delta s^{\rm e}_{kl},
\end{equation}
which guarantees that we can connect their ratios to the entropy flow $\Delta s^{\rm e}_{kl}$ into the environmental reservoir that mediates the transition~\cite{Seifert2012,VandenBroeck2015}.
For example, coupling to an auxiliary thermal bath at a different temperature $\beta^\prime$ entails $\Delta s^{\rm e}_{kl}=\beta^\prime(E_l-E_k)$ is proportional to the heat flux.
A biochemical example would be the conversion of ATP into ADP and $P_{\rm i}$ leading to $\Delta s^{\rm e}_{kl}=\beta (\mu_{ ATP}-\mu_{ ADP}-\mu_{ P_i})$, corresponding to the chemical work extracted from the ambient chemical baths.
It should be noted, however, that if such a chemical potential drop were erased, yet the system remained coupled to the baths, we would formally still include the ``driven" transitions that hydrolyze ATP in our representation, yet they would not occur at appreciable rates because they would lack the forward tilting provided by the favorability of conversation of ATP to ADP.  In order to eliminate such transitions from the picture completely, it would be necessary to take the system out of contact with the chemical baths. Yet, we should also point out that even in this ATP-free case, there would still be events involving passive catalysis by ATPase proteins that would in principle contribute to the undriven events represented in our Markov graph.

To characterize the minimum dissipation, we bound the total entropy production rate in the target nonequilibrium state.
As the system plus controller together is one open super-system with jump rates $\{{\mathcal R}_{ij}\}\equiv \{R_{ij},M_{ij}\}$, it must satisfy the second law of thermodynamics.
Namely, the entropy production rate must be positive~\cite{Seifert2012,VandenBroeck2015}:
\begin{equation}
{\dot S}_{\rm i}  =\sum_{i>j}J_{ij}(p)\ln \frac{{\mathcal R}_{ij}p_j}{{\mathcal R}_{ji}p_i}\ge 0,
\end{equation} 
which is typically split between the rate of change of the Shannon entropy $S(p)=-\sum_ip_i\ln p_i$, given as ${\dot S}(p)=\sum_{i>j}J_{ij}(p)\ln(p_j/p_i)$, and the entropy flow into the environment
${\dot S}_{\rm e}=\sum_{i>j}J_{ij}(p)\ln ({\mathcal R}_{ij}/{\mathcal R}_{ji})$.

Now, the super-system produces entropy in the steady state $p^*$ at a rate
\begin{equation}
{\dot S}_{\rm i}=\sum_{i>j}J_{ij}(p^*)\ln \frac{R_{ij}p^*_j}{ R_{ji}p^*_i}+\sum_{k>l}J_{kl}(p^*)\ln \frac{ M_{kl}p^*_l}{M_{lk}p^*_l}.
\end{equation}
Our goal is to find a lower bound on this sum, determined solely by the fixed system properties $\{R_{ij}\}$ and the target state $p^*$. The essential observation is that every control edge linking a pair of states contributes positively to the entropy production. 
Indeed, link-by-link we have~\cite{Horowitz2014}
 \begin{equation}
J_{kl}(p^*)\ln \frac{M_{kl} p^*_{l}}{M_{lk}p^*_{k}}=(M_{kl} p^*_{l}-M_{lk}p^*_{k})\ln \frac{M_{kl} p^*_{l}}{M_{lk}p^*_{k}}\ge 0,
\end{equation} 
since $x\ln x\ge \ln x$. The same link-wise positivity has also been shown as a consequence of a general fluctuation theorem for partial entropy production ~\cite{Shiraishi2014,Shiraishi2016}.  Thus, each control edge contributes superfluous dissipation, implying the only unavoidable dissipation occurs along the system's undriven links:
\begin{equation}\label{eq:edgeBound1}
{\dot S}_{\rm i}\ge {\dot S}_{\rm min} =\sum_{i>j} J_{ij}(p^*)\ln \frac{R_{ij} p^*_{j}}{R_{ji}p^*_{i}}\ge 0.
\end{equation}
No matter how control is implemented, the system inevitably jumps along the original links, and those on average dissipate irrecoverable energy into the environment when the system is in the target state $p^*$.

Physical insight into the factors regulating \eqref{eq:edgeBound1} is offered by using detailed balance \eqref{eq:db} to re-express \eqref{eq:edgeBound1} in terms of the nonequilibrium ratio $p^*/p^{\rm eq}$,
\begin{equation}
{\dot S}_{\rm min}=\sum_{i>j}R_{ij}p^{\rm eq}_{j}\left(\frac{p^*_{j}}{p^{\rm eq}_{j}}-\frac{p^*_{i}}{p^{\rm eq}_{i}}\right)\left(\ln\frac{p^*_{j}}{p^{\rm eq}_{j}}-\ln\-\frac{p^*_{i}}{p^{\rm eq}_{i}}\right).
\end{equation}
This formulation emphasizes that the minimum cost depends on two factors.  
First, it depends on how structurally different $p^*$ is from $p^{\rm eq}$: the further $p^*$ is from equilibrium the more dissipation required.
Second, the timescale is completely specified by the equilibrium dynamics through $R_{ij}p_{j}^{\rm eq}$: to push a system into a nonequilibrium state one must overcome the natural evolution of the system.

These observations can be made quantitatively precise by reformulating \eqref{eq:edgeBound1} using the information-theoretic relative entropy.
The relative entropy between two densities $f_i$ and $g_i$,  $D(f||g)=\sum_i f_i \ln f_i/g_i$, is an information-theoretic measure of distinguishability~\cite{Cover}.
In thermodynamics, the rate of decrease in relative entropy of a relaxing distribution $p(t)$ against the equilibrium state, $D(p(t)||p^{\rm eq})$, quantifies the dissipation via $-\partial_t D(p(t)||p^{\rm eq})=\sum_{i>j}J_{ij}(p)\ln(R_{ij}p_j/R_{ji}p_i)$~\cite{Esposito2011}, due to detailed balance \eqref{eq:db}.
From this observation, we recognize ${\dot S}_{\rm min}$ \eqref{eq:edgeBound1} as the entropy production rate we would observe in the instant $p^*$ begins to relax under the undriven dynamics:
\begin{equation}\label{eq:edgeBound2}
 {\dot S}_{\rm min}= -\partial_t^{\rm eq}D(p(t)||p^{\rm eq})\Big |_{p(t)=p^*},
 \end{equation}
where the notation $\partial_t^{\rm eq}$ emphasizes that the evolution is under the equilibrium dynamics. 
Equation \eqref{eq:edgeBound2}  quantifies the intuitive fact that it costs more to control a system the farther it is from equilibrium and the faster the equilibrium relaxation dynamics.
Notably, the relative entropy has recently been shown to emerge naturally in the energetic cost of self-assembly as well~\cite{Nguyen2016}. 

An important special case of our bound is isothermal control -- where the driven control transitions exchange heat with one thermal reservoir at inverse temperature $\beta$, as in our introductory two-state example.
For isothermal control, ${\dot S}_{\rm i}=\beta {\dot W}$ is the external work provided by the control, be it mechanical or chemical.
In addition, by introducing the nonequilibrium free energy ${\mathcal F}(p)=\langle E\rangle_p-TS(p)=F^{\rm eq}+D(p||p^{\rm eq})$~\cite{Esposito2011,Deffner2012}, our bound \eqref{eq:edgeBound2} simplifies to ${\dot W}\ge -\partial_t^{\rm eq}{\mathcal F}(p^*)$.
The controller must supply work at a rate that compensates the loss of free energy as the system tries to relax to equilibrium. 
This variant is reminiscent of a prediction for the minimum cost to control a quantum mesoscopic device~\cite{Horowitz2015b}, but that result is limited to control by an auxiliary feedback device.

\begin{figure}
\includegraphics[scale=.36]{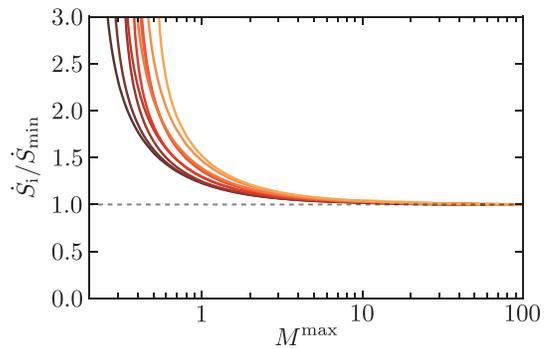}
\caption{Numerical verification of minimum dissipation bound: Completely connected graphs with $N=6$ nodes were randomly generated (different colors) with undriven rates such that for two states with $E_j>E_i$, $R_{ij}=e^{-B_{ij}}$ and $R_{ji}=e^{-(B_{ij}+E_j-E_i)}$, with barriers $B$ drawn from an unit-mean exponential distribution and energies from a zero-mean, unit-variance Gaussian distribution.
All possible driven edges were included. ${\dot S}_{\rm i}$ was numerically minimized subject to the constraints that $0< M_{ij}<M^{\rm max}$ and $p^*=1/6$ is uniform.
For fast enough $M^{\rm max}$, the bound \eqref{eq:edgeBound1} is saturated.}
\label{fig:numBound}
\end{figure}

Finally, our analysis readily offers the condition under which we saturate the minimum.
We reach the minimum dissipation when extraneous entropy production due to the controlled transitions is zero, {\it i.e.}~$J_{kl}(p^*)\ln (M_{kl}p^*_l/M_{lk}p^*_k)=0$. Thus, the optimal control rates $\{M_{kl}^*\}$ must verify
\begin{equation}
M_{kl}^*p^*_l=M_{lk}^*p_k^*.
\end{equation}
We can satisfy this condition only when the added edges operate much faster than the equilibrium transitions, guaranteeing that the controlled transitions are reversible.
In other words, fast control is optimal.
We verify this observation in Fig.~\ref{fig:numBound} by numerically minimizing ${\dot S}_{\rm i}$ for a random set of completely connected $N=6$ graphs.

\section{Implications}
Having formulated the general framework, we can immediately appreciate implications for various molecular processes of maintenance and self-repair.

\subsection{Molecular repair}

\begin{figure}[t]
\includegraphics[scale=.36]{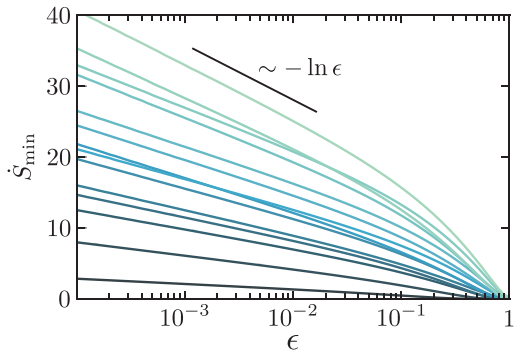}
\caption{Numerical verification of high fidelity control: Random completely connected graphs with $N=6$ nodes were generated as in Fig.~\ref{fig:numBound} with different realizations distinguished by color. ${\dot S}_{\rm min}$ with target distribution confined to one state with probability $1-\epsilon$ is plotted as function of the fidelity $\epsilon$, displaying a $-\ln \epsilon$ scaling.}
\label{fig:epsilon}
\end{figure}

First, consider a system with $N$ mesostates indexed by $i$, where we have the functional goal of ensuring that the system is found in a prescribed state, say $i=0$, with high probability $p_{0} =1-\epsilon$, where $\epsilon\ll 1$ is a small number that controls fidelity.  
Scenarios such as this are commonplace in biochemistry; in the cell, it is frequently the case that a chemical fuel such as ATP is used to pay for quality control in essential processes such as protein folding, nucleic acid replication, or polypeptide translation and degradation \cite{murugan2012speed,jiang2014reca,kim2013molecular}.
Here, the bound \eqref{eq:edgeBound1} predicts a $\epsilon$-scaling ${\dot S}_{\rm min}\sim -(1/\tau)\ln\epsilon$, where $\tau=(\sum_{j\neq0}R_{j0})^{-1}$ is the exit timescale from the target state, which we verify numerically in Fig.~\ref{fig:epsilon}. 
This scaling is consistent with the simple two-state model of chaperone action considered earlier: the limit $\epsilon\ll1$ implies that the dominant cost comes from maintaining fidelity and is insensitive to the background energy landscape.
It further matches well with past thermodynamic bounds derived specifically for biochemical error-correction~\cite{sartori2015thermodynamics}.

\subsection{Cytosolic protein localization}

As a final example, consider the cost of maintaining cytosolic protein localization. Recent studies using fluorescence microscopy in eukaryotic cells have revealed a wide range of diffusively open sub-cellular compartments not enclosed by membranes, which coalesce or disassemble rapidly under cellular stresses, such as nutrient starvation or heat shock~\cite{kaganovich2008misfolded,narayanaswamy2009widespread,brangwynne2009germline}. While evidence in particular cases suggests the formation of such structures could be an equilibrium phase separation \cite{li2012phase}, it is possible in principle that the cell exploits nonequilibrium driving to a maintain spatial order  without employing attractive interparticle interactions that retard diffusive mobility  \cite{brangwynne2015polymer}.

As a simple model of this situation, consider a solution of $N$ proteins composed of two chemical species $A$ and $B$ diffusing in a region $V$ with equal diffusivities $D$.  
We wish to confine all of the $A$ proteins, numbering $N_A=fN$, to a region $\nu$, while displacing $B$ proteins, thereby maintaining a uniform total concentration. Although the bound derived above also applies in far more general scenarios, we will assume the chemical monomers are non-interacting for the sake of calculational simplicity.

The minimum dissipation rate in this diffusive limit is obtained by first imagining we have a single molecule making a random walk on a $d$-dimensional square lattice with equal transition rates $k$, implying a uniform energy landscape.
We then shrink the lattice spacing as $\Delta x\to 0$, while diffusively accelerating time $k\to D/(\Delta x)^2$, allowing us to approximate \eqref{eq:edgeBound1} as
\begin{align} \label{eq:entropy_functional}
{\dot S}_{\rm min}&= \sum_{\langle {\bf i}{\bf i}^\prime\rangle} k(p^*_{{\bf i}^\prime}-p^*_{\bf i})\ln\frac{p^*_{{\bf i}^\prime}}{p^*_{{\bf i}}}\\
&\approx D\int \frac{\nabla p^*({\bf x})\cdot \nabla p^*({\bf x})}{p^*({\bf x})}\,  d{\bf x},
\end{align}
where the summation is over pairs of neighboring lattice sites.
Confining $A$ to $\nu$ under the constraint that the total concentration is constant $fp_A({\bf x})+(1-f)p_B({\bf x})=1$ -- where $p_j({\bf x})$ with $j=A,B$ is the probability density of species $j$ to be found at location ${\bf x}$ -- suggests an ultimate minimum cost
\begin{equation} \label{eq:entropy_functional}
{\dot S}_{\rm min}=\min_{p_A^*\in \nu} \sum_{j=A,B}N_jD\int \frac{\nabla p^*_j({\bf x})\cdot \nabla p^*_j({\bf x})}{p^*_j({\bf x})}\, d{\bf x},
\end{equation}
assuming independent $A$ and $B$.

Assuming a cytosolic mass density of $300\, {\rm mg}/{\rm mL}$ \cite{luby1999cytoarchitecture} filled with $25\, {\rm kDa}$ globular proteins, the confinement of a single protein to a cubic region $\nu$ of side $L=1\, \mu{\rm m}$, corresponds to a choice of $f\simeq 10^{-7}$, so we can stipulate that $f=N_A/N\ll1$.
In this limit, the optimal distribution of $B$ molecules $p^*_B$ is uniform, whereas the optimal distribution of $A$ molecules is 
\begin{equation}
p^*_A({\bf x})=\prod_i[1-\cos(2\pi x_i/L)]/L.
\end{equation}
The resulting minimum work cost per confined protein at physiological temperature $T$ is $\dot{W}/fN=T{\dot S}_{\rm min}/fN=3k_{B}TD\left(2\pi/L\right)^2$.  For a diffusion coefficient of a small globular protein like GFP, for which $D = 26\, {\mu\rm{m}}^2/{\rm s}$, the predicted number of ATP hydrolyzed per confined protein is roughly $10^2\, {\rm molecules}/{\rm s}$~\cite{Alberts}.  Notably, this rate is larger by a factor of $\sim 1-100$ than the rate of heat dissipation per protein in exponentially growing microbes \cite{England2013}.  This comparison suggests that the energetic cost of nonequilibrium confinement could significantly impact when and how the cell might benefit from such a mechanism.

JMH and JLE are supported by the Gordon and Betty Moore Foundation through Grant GBMF4343. JLE further acknowledges the Cabot family for their generous support of MIT.

\bibliography{PhysicsTexts.bib,FluctuationTheory.bib,mybib.bib} 

\end{document}